\documentclass{kluwer}    
\usepackage{graphicx}                                                                                                       
\newdisplay{guess}{Conjecture}

\begin{document}                                                                                   
\begin{article}
\begin{opening}         
\title{Searching for Line Profile Variability in HgMn Stars}
\author{Sylvain \surname{Turcotte}}  
\runningauthor{Turcotte et. al}
\runningtitle{LPVs in HgMn Stars}
\institute{Lawrence Livermore National Laboratory, L-413, P.O. Box 808,
Livermore, CA 94551, USA}
\author{Conny \surname{Aerts}}  
\institute{Instituut voor Sterrenkunde, Katholieke Universiteit Leuven,
Celestijnenlaan 200 B, B - 3001 Leuven, Belgium}
\author{Paul \surname{Knoglinger}}  
\institute{Institute for Astronomy, University of Vienna,
Turkenschanzstrasse 17, A-1180 Vienna, Austria}
\date{August 31, 2002}

\begin{abstract}
Spectra of four non-magnetic chemically peculiar late B type stars (HgMn) stars 
are analysed to detect periodic spectral line
variations (LPVs). A procedure developed to study LPVs in Slowly Pulsating B stars
has been adopted as pulsational properties of HgMn stars should be expected to be
similar. In the preliminary results discussed here no conclusive evidence for
periodic LPVs was uncovered. A more sensitive re-analysis of the data is
under way.
\end{abstract} 
\keywords{sample, \LaTeX}

\end{opening}           

\section{Motivation}  

HgMn stars are chemically peculiar stars for which periodic variability
has not been found as of yet. Searches for variability have been made
mostly photometrically though some studies of spectral variability have
also been attempted.

Historically, several HgMn stars have been claimed to be variable
but variability as yet to be confirmed in any of them
\cite{Adelman98}. A large number of HgMn stars were observed as part of the
Hipparcos mission but no periodic variability was detected. The maximum
permitted amplitude can in many cases be expected to be at most a few mmag.
Recently, some spectral variability was claimed in
$\alpha$~Andromed{\ae} which were
interpreted as possible surface chemical inhomogeneities
\cite{Adelmanetal02}. The authors argued that such variability would be
the exception rather than the rule in HgMn stars.

The pursuit of elusive evidence of variability, both spectroscopically
and photometrically, is motivated by several unresolved questions:
\begin{itemize}
 \item pulsations is expected theoretically from current models, in
   other words confirmation of stability or the discovery of low amplitude
   pulsations can provide constraints on physical processes not accounted
   for in the models (see Turcotte \& Richard in these proceedings);
 \item rotational variability would provide evidence of surface
   inhomogeneities related to diffusion, mass loss and/or magnetism in 
   the atmosphere of B stars;
 \item confirm or infirm that all HgMn stars are part of binary or
   multiple systems which could help answer the question as to how B stars
   can be slowly rotating in the absence of binarity or magnetism.
\end{itemize}

In this short paper we present preliminary results of the search of 
line profile variability in a substantial series of echelle spectra 
of four bright HgMn stars of the southern hemisphere. These observations
represent an unprecedented effort to study spectroscopic variability in
HgMn stars and are expected to help put stronger constraints on 
pulsations in these stars.

\section{The Program Stars}

\begin{figure} 
\centerline{\includegraphics[width=12pc]{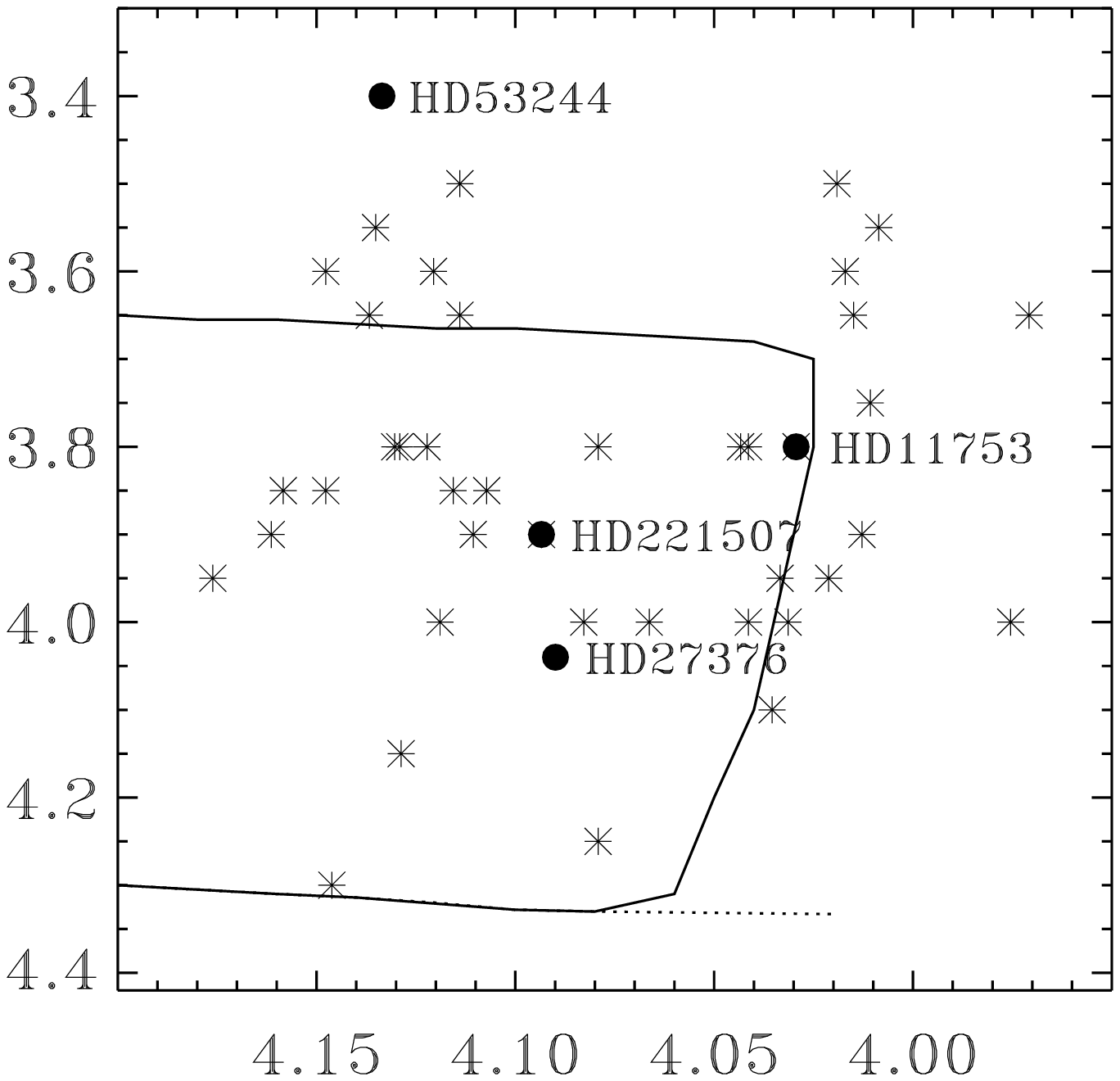}}
\caption[]{$\log g$-T$_{\rm eff}$ diagram showing the program stars and the
           theoretical limit of the SPB instability region
           \cite{Pamyatnykh99} along with
           a sample of other HgMn stars \cite{SmithDworetsky93}.
}
\label{fig:HRD}
\end{figure}

The four program stars were the brightest southern HgMn stars 
visible during the periods of observation (see next section). Three of
the four are within the theoretical instability region for SPB stars
(HD~11753 being right on the cool edge),
the fourth (HD~53244) being slightly too evolved (Figure~\ref{fig:HRD}). 

The spectra were taken over two campaigns of several days, from 
September 28$^{\rm th}$ to October 11th$^{\rm th}$ and from 
December 2$^{\rm nd}$ to December 15$^{\rm th}$ 2000, with the
CORALIE spectrograph at the 1.2~m telescope at La Silla. 
The observations are summarized in Table~\ref{tab:obs}.
\begin{table} %
\begin{tabular}{cccccc}
\hline
HD     & name         &  V   & \# nights & \# spectra & exposures (s) \\
11753  & $\phi$~Phe   & 5.11 & 26        & 105        & 600 to 1750      \\
27376  & 41~Eri       & 3.55 & 11        & 28         & 160 to 800       \\
53244  & $\gamma$~CMa & 4.10 & 20        & 74         & 230 to 800       \\
221507 & $\beta$~Scl  & 4.37 & 26        & 108        & 7 to 1200        \\
\hline
\end{tabular}
\caption[]{Summary of observations of the program stars}\label{tab:obs}
\end{table}

\begin{figure} 
\centerline{\includegraphics[width=20pc]{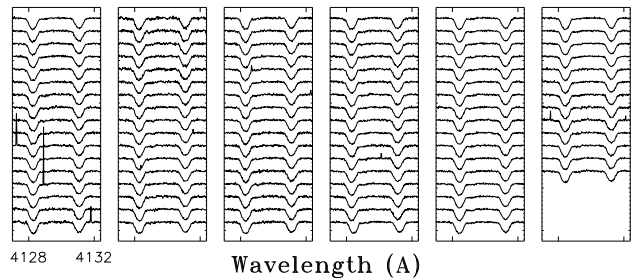}}
\caption[]{The SiII doublet for the set of spectra for HD221507 used in searching
for LPVs.
}
\label{fig:spec}
\end{figure}

Due to space constraints we henceforth discuss only the star for which the
better results were obtained at this point in the analysis, HD221507.
The spectra selected for this star after bad data was removed are shown
in Figure~\ref{fig:spec}.

\section{Preliminary Results}

We focused on the Si{II} doublet at $\lambda$4128.053 and
$\lambda$4130.884~{\AA} for which the first moment was calculated, a
procedure developed to study SPB stars \cite{DeCat01}.
The variability was studied using the PDM method.
The models of HgMn stars suggest that they should pulsate in a similar
way to SPB stars, if at all. 

\begin{figure} 
\centerline{\includegraphics[width=15pc]{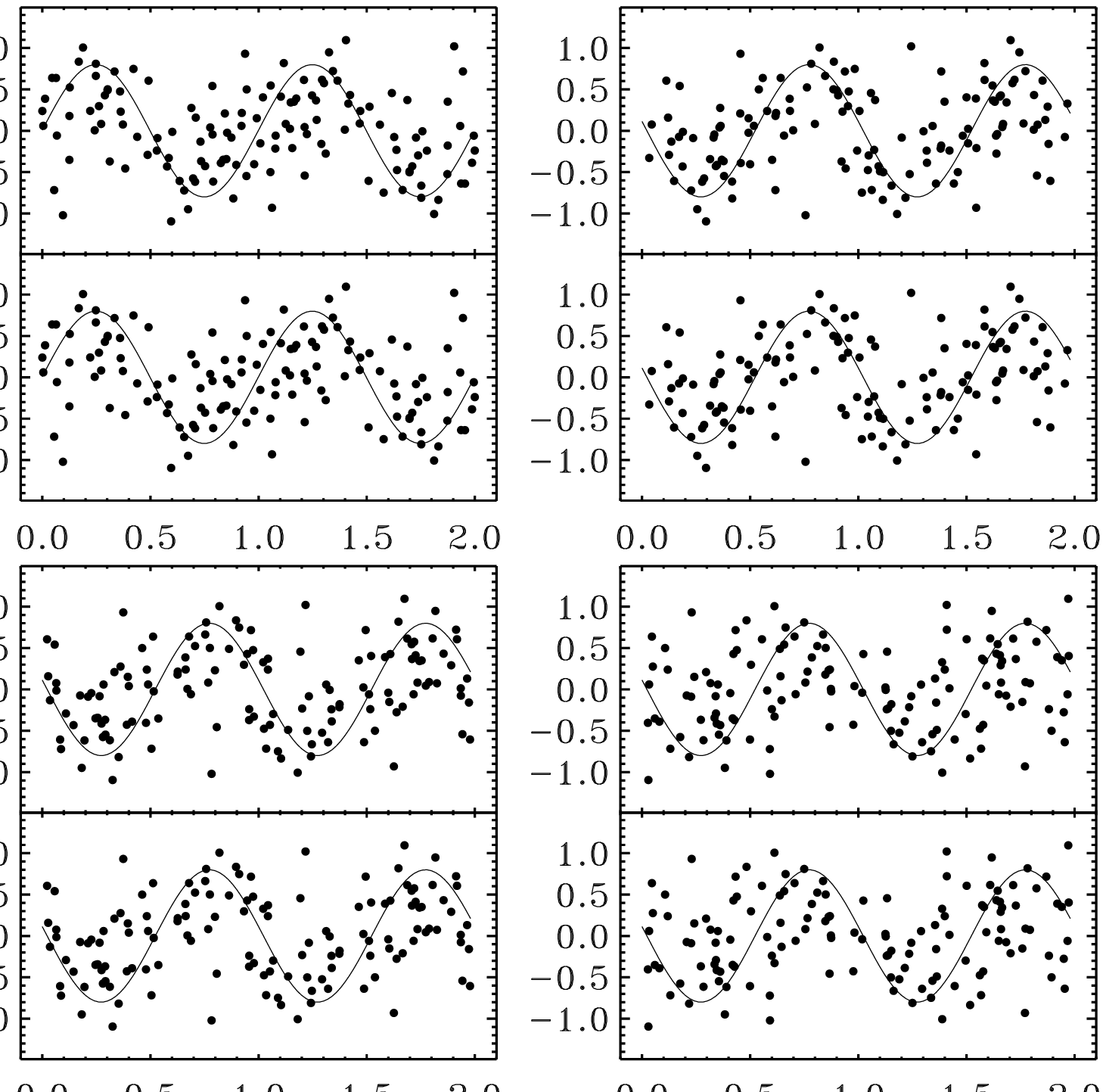}}
\vskip 8pt
\caption[]{Phase plots for the four {\sl best} periods, 0.31, 0.44,
0.78, 1.38~$d$ clockwise starting from the upper left, for HD221507
with a sine wave eye fitted for amplitude and phase. 
}
\label{fig:phase}
\end{figure}

Four phase plots are shown in Figure~\ref{fig:phase}. The periods
shown, 0.31, 0.44, 0.78, 1.38~$d$ were the ones which would reproduce 
the best approximation  to a sine wave. The periods are in the range
expected for SPBs. The scatter is evidently quite large in all cases 
and the variability, although somewhat suggestive, is far from clear.

The data analysis used to obtain the preceding results was not refined
and we expect that significantly improved sensitivity will be achieved
with forthcoming work on these data.

\acknowledgements
This work was performed in part under the auspices of the U.S.
Department of Energy, National Nuclear Security Administration by the
University of California, Lawrence Livermore National Laboratory under
contract No.W-7405-Eng-48.

\end{article}
\end{document}